\documentclass[onecolumn]{revtex4-2}
\usepackage[utf8]{inputenc}
\usepackage[T1]{fontenc}
\usepackage[]{graphicx}
\usepackage{grffile}
\usepackage{longtable}
\usepackage{wrapfig}
\usepackage{rotating}
\usepackage[normalem]{ulem}
\usepackage{amsmath}
\usepackage{textcomp}
\usepackage{amssymb}
\usepackage{capt-of}
\usepackage{hyperref}
\usepackage{natbib}
\usepackage[margin=1in]{geometry}
\setcitestyle{square}
\usepackage{color}
\usepackage[breakable,most,many]{tcolorbox}
\usepackage{ulem}

\begin{document}

\title{Robust retrieval of dynamic sequences through interaction modulation}

\author{Lukas Herron}
\affiliation{Biophysics Program and Institute for Physical Science and Technology, University of Maryland, College Park, MD, United States}

\author{Pablo Sartori}
\affiliation{Instituto Gulbenkian de Ci\^encia, Oeiras, Portugal}

\author{BingKan Xue*}
\affiliation{Department of Physics, University of Florida, Gainesville, FL, United States}

\begin{abstract}
Many biological systems dynamically rearrange their components through a sequence of configurations in order to perform their functions. Such dynamic processes have been studied using network models that sequentially retrieve a set of stored patterns. Previous models of sequential retrieval belong to a general class in which the components of the system are controlled by a feedback (``input modulation''). In contrast, we introduce a new class of models in which the feedback modifies the interactions among the components (``interaction modulation''). We show that interaction modulation models are not only capable of retrieving dynamic sequences, but they do so more robustly than input modulation models. In particular, we find that modulation of symmetric interactions allows retrieval of patterns with different activity levels and has a much larger dynamic capacity. Our results suggest that interaction modulation may be a common principle underlying biological systems that show complex collective dynamics.
\end{abstract}

\maketitle

\section{Introduction}

\begin{figure}%
    \centering
    \includegraphics[
    ]{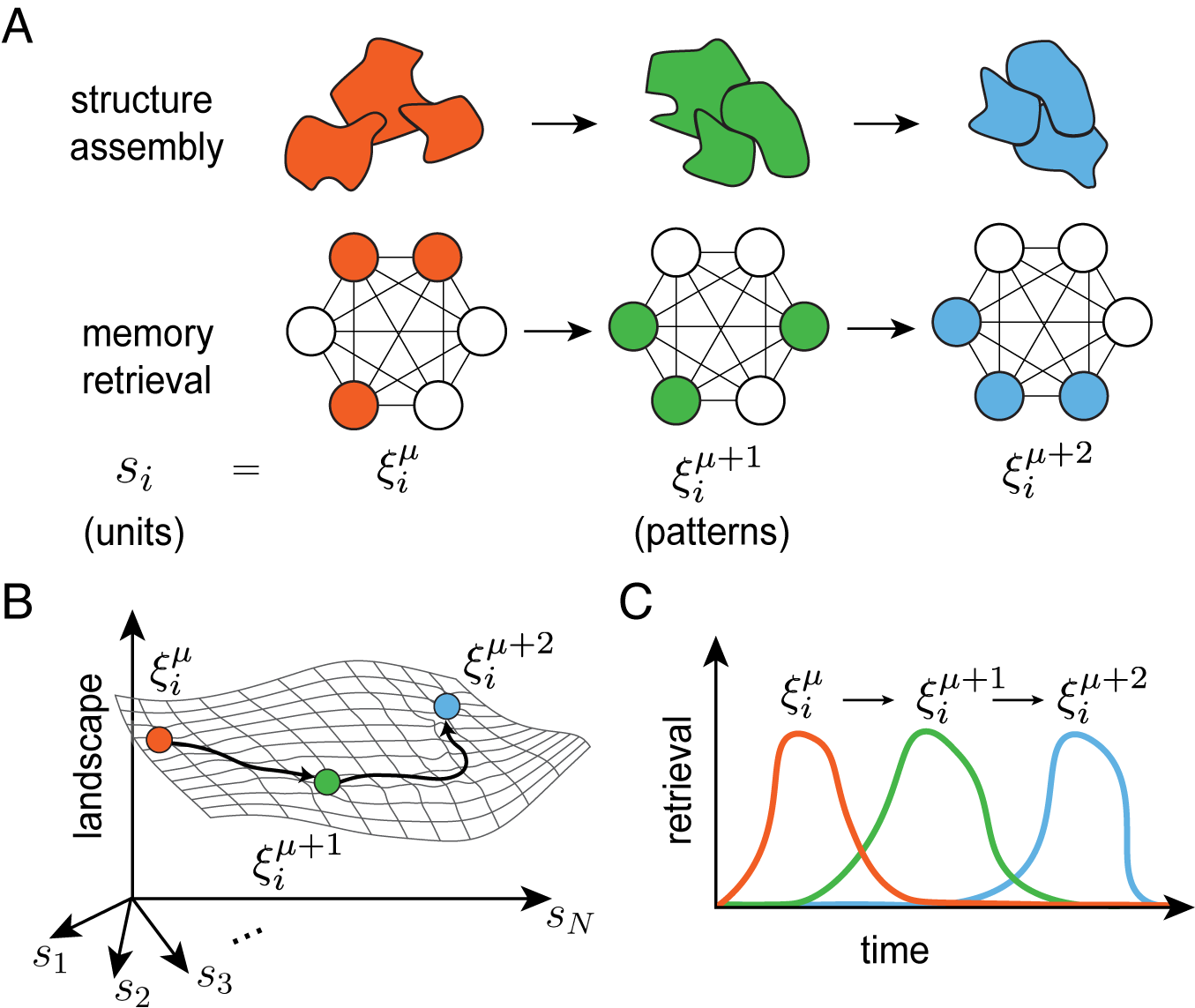}
    \caption{{\bf Sequential dynamics in complex systems.} {\bf A}: Sequential transition along a sequence of configurations is a common phenomenon in complex systems, such as the assembly of multi-protein complexes and the synaptic activity of neurons. Such dynamics may be modeled by a network where the units $s_i$ store multiple patterns $\xi^\mu$. {\bf B}:  Sequential dynamics can be viewed as unfolding across a rugged, changing landscape. {\bf C}: Transitions between configurations are represented by time series of order parameters, which measure how close the system is to each configuration. The peak of each colored curve represents the retrieval of a particular configuration.}
    \label{fig:schematic-seqdyn}
\end{figure}

Biological systems are often made of many components that dynamically arrange themselves into specific configurations. In some cases, this dynamics follows a particular sequence of configurations which allows the system to perform some function, as illustrated in Fig.~\ref{fig:schematic-seqdyn}. For example, neurons rearrange their synaptic activity in order to generate a sequence of activity patterns \cite{wallenstein1998hippocampus, Eichenbaum_2013}. Multi-protein assemblies dynamically rearrange their protein composition several times in a precise order as they perform a function, such as the spliceosome processing pre-mRNAs \cite{wahl2009spliceosome}. Bacterial communities on marine particles undergo successions where the species structure changes in reproducible patterns to degrade organic matter \cite{Datta2016}. Thus, sequential transitions of multi-component systems through well-defined configurations is a general phenomenon in biology. 

These specific configurations can be considered metastable states of the dynamics. The ability of the system to transition from one such configuration to the next means that the system can alter the stability of the configurations. This can be achieved by controlling specific components, e.g., by changing the abundance of certain proteins in the case of assembly dynamics, or by modulating the neuron firing activity in neural systems. Such a scenario corresponds to a common approach in control theory, i.e., modulating \textit{inputs} on a subset of variables to influence the full system \cite{Liu2016}. In biology, however, a different approach may be considered, namely to modify the \textit{interactions} among the components. Unlike physical systems where the interactions between elementary particles are determined by fundamental forces, biological components are complex objects and the effective interaction strengths between them can be modified by third parties. For example, the affinity between a pair of proteins can be regulated by allosteric modulation through a third protein ~\cite{Wodak2019}. Similarly, the synapses among a pair of neurons can be modulated by a third neuron via heterosynapstic plasticity \cite{Bailey2000}. Can such interaction modulation be used to control the dynamics of complex systems? What are the differences in performance between interaction and input modulation?

In this work we address these questions using a framework inspired by the Hopfield neural network \cite{Hopfield1982}. The Hopfield network was originally developed as an abstract model of associative memory capable of storing and retrieving particular network configurations. This paradigm has been extended to model biological systems ranging from metabolic networks \cite{de2012reaction} to protein assemblies \cite{sartori2020lessons} and even ecosystems \cite{Power2015}. Furthermore, sequential transitions among the stored configurations have long been considered \cite{sompolinsky1986temporal, horn1989neural, dehaene1987neural, buhmann1987noise}, aimed at describing phenomena such as central pattern generation \cite{kleinfeld1988associative}, counting \cite{amit1988neural} and, more recently, free association \cite{russo2012cortical}, memory recall \cite{naim2020fundamental}, and assembly dynamics \cite{osat2022non}. Here, we model the dynamics of retrieval by introducing a small set of feedback units, which control the sequential transitions. When formulated this way, previous models are shown to fall into a class of models based on \emph{input modulation}. We propose a new class of models that rely on \emph{interaction modulation} to trigger autonomous transitions between configurations. Remarkably, we find that modulation of symmetric interactions allows sequential retrieval of configurations that have different activity levels, which cannot be reliably done by models that use input modulation. Furthermore, this model can retrieve much longer sequences than other models. Our results suggest that interaction modulation may be biologically favored over input modulation due to its robustness and large dynamic capacity.

\section{Pattern storage and retrieval} \label{sec:framework}

Our model is based on the classic Hopfield network, which can store and retrieve a given set of network configurations, or ``patterns''. The network is composed of $N$ units, whose activities are denoted by $\{s_i\}_{i=1}^N$ and take continuous values, $0 \leq s_i \leq 1$. The interactions among the units are characterized by a matrix $W_{ij}$, which contributes to the input of each unit:
\begin{align}
\label{eq:neural-inputs}
  h_i = \sum_{j=1}^N W_{ij} s_j + V_i  \quad.
\end{align}
Here $V_i$ is an external input to each unit that does not depend on the current state $s_i$ of the network. The dynamics of the system is governed by
\begin{align}
    \label{eq:dyn-sys}
    \dot{s}_i = -s_i + F(h_i)\quad,
\end{align}
where $F(\cdot)$ is an activation function centered at zero, such as the sigmoid $F(x) = 1 / (1 + \mathrm{e}^{-\beta x})$.

The original task of a Hopfield network is to store and retrieve $p$ patterns, denoted by $\{{\xi}^\mu\}_{\mu=1}^p$, each of which is a vector with binary elements, ${\xi}^\mu_i \in \{0,1\}$.  The patterns can represent structured content, such as the pixels of black-and-white images, but for simplicity we take them to be random with an average activity $a$. That is, each pattern has a fraction $a$ of the units set to $1$ and the rest set to $0$. To store those patterns, the following symmetric interactions are introduced \cite{Hopfield1982},
\begin{align}
    \label{eq:hopfield-sym}
    J_{ij} &= \frac{1}{N} \sum_{\mu=1}^p (\xi_i^\mu - a) (\xi_j^\mu - a) \quad.
\end{align}
For $W_{ij} = J_{ij}$ and $V_i = 0$, the patterns are fixed points of the dynamics in Eq.~(\ref{eq:dyn-sys}). That is, the system will retrieve a pattern provided that it is close to it initially, where the proximity to patterns is measured by the $p$ overlap variables,
\begin{equation} \label{eq:overlap}
m^\mu = \frac{1}{Na(1-a)} \sum_i (\xi_i^\mu - a) s_i \quad.
\end{equation}
Thus, for $s_i = \xi^{1}_i$, we have $m^{1} = 1$ and $m^{\mu\neq1} \approx 0$, because random patterns are approximately orthogonal for large $N$. Eqs.~(\ref{eq:neural-inputs}-\ref{eq:hopfield-sym}) define a dynamical system capable of storing and retrieving each individual pattern.

We are interested in networks that can autonomously retrieve a sequence of patterns, one after another. Already in Hopfield's original paper \cite{Hopfield1982}, it was suggested that sequential retrieval could be achieved by introducing asymmetric interactions of the form
\begin{align} \label{eq:hopfield-asym}
    \tilde{J}_{ij} &= \frac{1}{N} \sum_\mu (\xi_i^{\mu+1} - a) (\xi_j^\mu - a) \quad.
\end{align}
These asymmetric interactions provide a directional bias from every pattern $\xi^\mu$ towards the next pattern $\xi^{\mu+1}$. The interaction matrix is then generalized to $W_{ij} = J_{ij} + \lambda \tilde{J}_{ij}$, where the parameter $\lambda$ represents the strength of the directional bias. The rationale behind this construction is that, after a pattern is retrieved due to the symmetric $J_{ij}$ term, the asymmetric $\tilde{J}_{ij}$ term will \textit{destabilize} it and push the system toward the next pattern in the sequence, i.e., $\xi^1 \to \xi^2 \to \xi^3 \cdots$.

However, this simple model cannot produce sequences reliably \cite{Hopfield1982}. Instead, the network exhibits no dynamics for $\lambda$ less than a certain value, or chaotic dynamics otherwise \cite{sompolinsky1988chaos}. The reason is that the $J_{ij}$ and $\tilde{J}_{ij}$ terms act on the same timescale, such that either the stabilizing term dominates and leads to no dynamics, or the destabilizing term dominates and leads to chaotic dynamics. Therefore, robust sequential dynamics requires a separation of timescales between fast stabilization and slow destabilization. In this way, the system first relaxes to a pattern $\xi^\mu$, which is slowly destabilized, then goes to the next pattern $\xi^{\mu+1}$, and so on.

\begin{figure}%
    \centering
    \includegraphics[
    ]{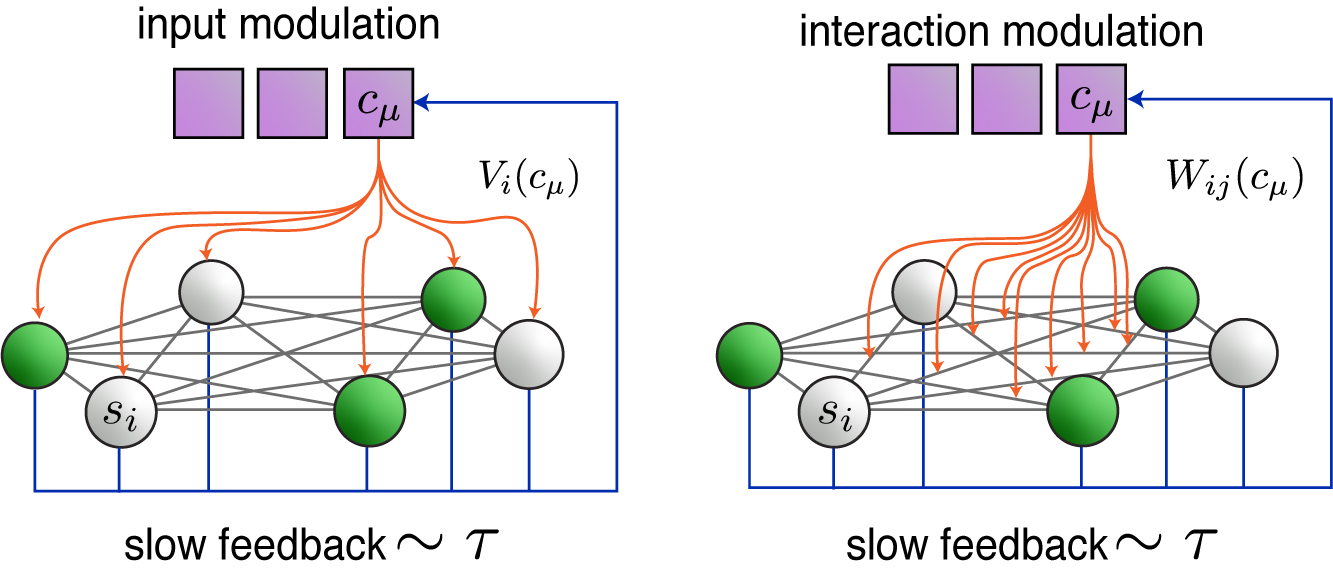}
    \caption{{\bf Input versus interaction modulation.} A Hopfield network with units $s_i$ (spherical nodes) is controlled by a set of feedback units $c_\mu$ (square nodes). These feedback units are updated on a slow timescale $\tau$ (blue lines), and modulate either the input field $V_i$ of the main units or their interactions $W_{ij}$ (orange lines).}
    \label{fig:schematic-modulation}%
\end{figure}

The required separation of timescales can be achieved through feedback modulation. To this end, we introduce a set of feedback units represented by the variables $\{ c_\mu \}_{\mu=1}^p$, which obey the dynamics
\begin{align}
    \label{eq:c-mu-ODE}
    \dot{c}_\mu = -\frac{1}{\tau} (c_\mu - m^\mu) \quad.
\end{align}
These feedback units will be used to destabilize the retrieved pattern on a timescale $\tau \gg 1$, by either modifying the inputs to the network, i.e., $V_i(c_\mu)$, or the interactions, $W_{ij}(c_\mu)$, as schematically depicted in Fig.~\ref{fig:schematic-modulation}. We will elaborate on these two approaches below, with four representative models summarized in Box 1.

\begin{tcolorbox}[
        float*=t,
 		colback=white,
		boxrule=0.7pt,
		sharp corners,
		parbox=false,
		width=\textwidth,
		lower separated=false,
		title=Box 1 --- Models of sequential retrieval
	]

Sequential retrieval can be achieved via feedback control that modulates either the input to the units or the interactions between the units. Two classical models of sequential retrieval via input modulation (HU and SK) and two new models via interaction modulation (MSI and MAI) are summarized in the table below, and elaborated on in Section \ref{sec:interaction-input}.

\begin{center}
\setlength{\tabcolsep}{6pt}
\renewcommand{\arraystretch}{1.8}
\begin{tabular}{|c|c|c|c|}
\hline
Model & Interaction $W_{ij}$ & Field $V_i$ & Feedback through $c_\mu$ \\
\hline
HU \cite{horn1989neural} & $J_{ij} + \lambda \, \tilde{J}_{ij}$ & $-\theta \, U_i(c_\mu)$ & $U_i(c_\mu) \equiv \sum_\mu \xi_i^\mu c_\mu$ \\
\hline
SK \cite{sompolinsky1986temporal} & $J_{ij}$ &  $\lambda \, {U}_i(c_\mu) - \theta$ & ${U}_i(c_\mu) \equiv \sum_\mu (\xi_i^{\mu+1} - a) c_\mu$ \\
\hline
MSI & $J_{ij}(c_\mu) + \lambda \, \tilde{J}_{ij}$ & $-\theta$ & $J_{ij}(c_\mu) \equiv \frac{1}{N} \sum_\mu (\xi_i^{\mu+1} - a) (\xi_j^{\mu+1} - a) c_\mu$ \\
\hline
MAI & $J_{ij} + \lambda \, \tilde{J}_{ij}(c_\mu)$ & $-\theta$ & $\tilde{J}_{ij}(c_\mu) \equiv \frac{1}{N} \sum_\mu (\xi_i^{\mu+1} - a) (\xi_j^{\mu} - a) c_\mu$ \\
\hline
\end{tabular}
\end{center}
The figures below provide a heuristic depiction of the dynamics for each model. The ``energy landscape'' represents the patterns stored in the network. The network state $s_i$ currently occupies the pattern $\xi^\mu$ and will transition to $\xi^{\mu+1}$. The black arrows represent update to the feedback units $c_\mu$ on a timescale $\tau$, and the orange lines represent the effect of feedback. The red arrows represent directional bias that guides the transitions, and the dashed gray lines represent the effect of input fields (constant when not depicted). The input modulation models work by raising the inhibitory threshold (HU) or tilting the energy landscape (SK), whereas the interaction modulation models work by enforcing the directional bias (MAI) or deforming the energy landscape (MSI) to modify the stability of the minima.
\begin{center}
\begin{minipage}[t!]{\linewidth}
\center
\includegraphics[]{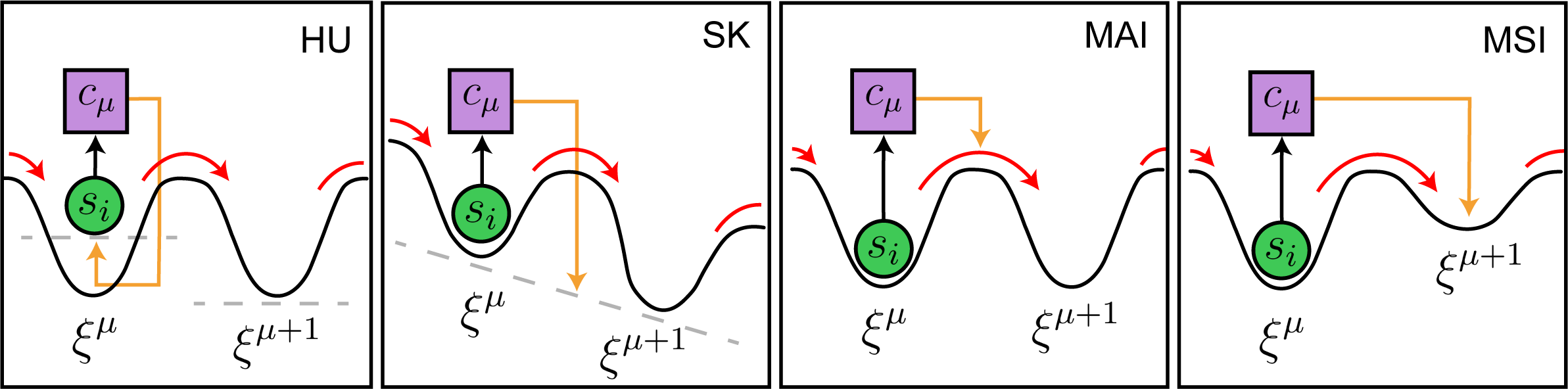}
\end{minipage}
\end{center}
\end{tcolorbox}

\section{Models of sequential retrieval} \label{sec:models}

\subsection{Sequential retrieval via input modulation} \label{sec:interaction-input}

We first reformulate some well-studied models of sequential retrieval using our framework. For instance, the model due to Horn \& Usher \cite{horn1989neural} can be re-expressed by setting $W_{ij} = J_{ij} + \lambda \tilde{J}_{ij}$ and $V_i = -\theta \sum_\mu \xi_i^\mu c^\mu$. The latter is usually referred to as an adaptive threshold, with $\theta$ as the strength of inhibition. In this model neurons that are active in a retrieved pattern will experience an inhibitory input after a time $\sim \tau$, which tends to turn them off and thus destabilize the current pattern. This model uses separate terms to serve different purposes: the symmetric interactions $J_{ij}$ stabilize each pattern, the external field $V_i$ slowly destabilizes the retrieved pattern through the feedback coupling, and the asymmetric interactions $\tilde{J}_{ij}$ bias the system towards the subsequent pattern. Fig.~\ref{fig:examples}A shows an example of sequential retrieval using this model.

Similarly, we can reformulate the model due to Sompolinsky \& Kanter \cite{sompolinsky1986temporal} through identifying $W_{ij} = J_{ij}$ and $V_i = \lambda \sum_\mu (\xi_i^{\mu+1} - a) c_\mu - \theta$. In this model, after the network retrieves a pattern, the feedback units $c_\mu$ slowly activate to drive the system towards the subsequent pattern. This is enough to destabilize the current pattern if $\lambda$ is sufficiently large (see Section~\ref{sec:phase} for feasible parameter regions). An example of this dynamics is shown in Fig.~\ref{fig:examples}B.

\subsection{Sequential retrieval via interaction modulation}

\begin{figure}
    \centering
    \includegraphics[
    ]{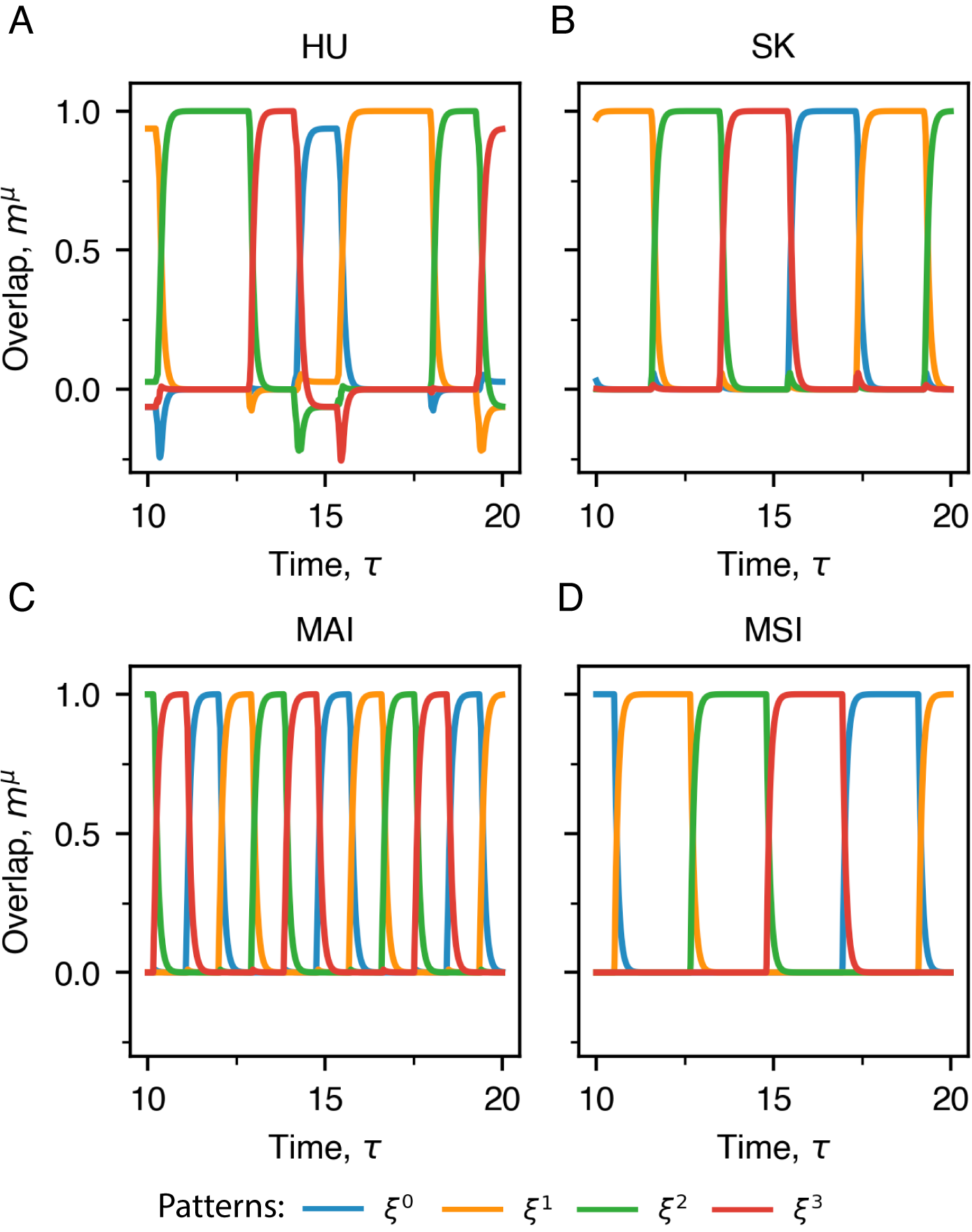}
    \caption{{\bf Examples of dynamic retrieval for input and interaction modulation models.} {\bf A}-{\bf B}: The HU and SK models that belong to the input modulation class. {\bf C}-{\bf D}: Modulation of asymmetric interactions (MAI) and symmetric interactions (MSI). Each model stores a cyclic sequence of four orthogonal patterns ($p=4$). Each color represents the overlap with a particular pattern $\xi^\mu$, which is retrieved when the overlap $m^\mu$ approaches 1. As different overlaps sequentially increase and decrease, the patterns are retrieved one after another, corresponding to retrieval of the stored sequence ($\xi^0 \rightarrow \xi^1 \rightarrow \xi^2 \rightarrow \cdots$). The parameters used for each model correspond to the green dots in Fig. \ref{fig:seqdyn-phasespace} (SK: $\lambda = 1.2, \theta = 0.37$; HU: $\lambda = 0.3, \theta = 0.62$; MAI: $\lambda = 1.7, \theta = 0.325$; MSI: $\lambda = 0.1, \theta = 0.06$).} 
    \label{fig:examples}
\end{figure}

Our reformulation of both models above makes it clear that in these models the feedback units $c_\mu$ modulate the input $V_i$ to achieve sequential retrieval. We now present new models of sequential retrieval in which the feedback units $c_\mu$ modulate the interactions $W_{ij}$. Two types of such modulation are possible, which act either on the symmetric interactions $J_{ij}$ or on the asymmetric interactions $\tilde{J}_{ij}$.

First consider a model for the modulation of the symmetric interactions (MSI), which can be described by $J_{ij} = \frac{1}{N} \sum_\mu (\xi_i^{\mu+1} - a) (\xi_j^{\mu+1} - a) c_\mu$. When the network has retrieved a pattern $\xi^\nu$ for a period of time, all $c_\mu$ will decay to zero except $c_\nu$. As a consequence, all terms in $J_{ij}$ will be turned off except for $\mu=\nu$. Therefore, only one pattern $\xi^{\nu+1}$ will be stable, which is the one that the network will retrieve subsequently. In other words, the symmetric interactions store and retrieve one pattern at a time. Asymmetric interactions $\tilde{J}_{ij}$ are needed to provide directional bias towards subsequent patterns, even though the strength $\lambda$ can be small (see Section~\ref{sec:phase}).  

Alternatively, feedback units can be used to modulate the asymmetric interactions (MAI). Consider a model with $\tilde{J}_{ij} = \frac{1}{N} \sum_\mu (\xi_i^{\mu+1} - a)(\xi_j^{\mu} - a) c_\mu$. As before, when the network has retrieved a pattern $\xi^\nu$ for some time, $c_\nu$ reaches a large value while all other $c_\mu$ decay to zero. In this model, however, only one directional bias is active, corresponding to the transition $\xi^\nu\to\xi^{\nu+1}$. For a sufficiently large $\lambda$, this term will destabilize the current pattern and push the system towards $\xi^{\nu+1}$. Compared to the MSI model above, here all patterns are stored in the symmetric interactions, but only one transition is enabled at a time.

In Fig.~\ref{fig:examples}C and D we show two examples of sequential retrieval using the MSI and MAI models, respectively. As one can see, the dynamic trajectories are very similar to the HU and SK models, which operate via input modulation. We therefore conclude that interaction modulation is an equally feasible way of retrieving dynamic sequences.

\section{The phase space of sequential dynamics} \label{sec:phase}

Within our general framework, all models of sequential retrieval are characterized by the same two parameters: the magnitude of the bias $\lambda$ and the threshold $\theta$. This allows us to compare interaction and input modulation by systematically examining the $(\lambda,\theta)$ parameter space and identifying the regions in which sequential retrieval occurs. To proceed, we numerically solved the dynamical system and used a custom-made score for the dynamics to quantify the accuracy of sequential retrieval, see Appendix~\ref{sec:computational-details} for details.

\begin{figure*}%
    \centering
    \includegraphics[
    width=\textwidth
    ]{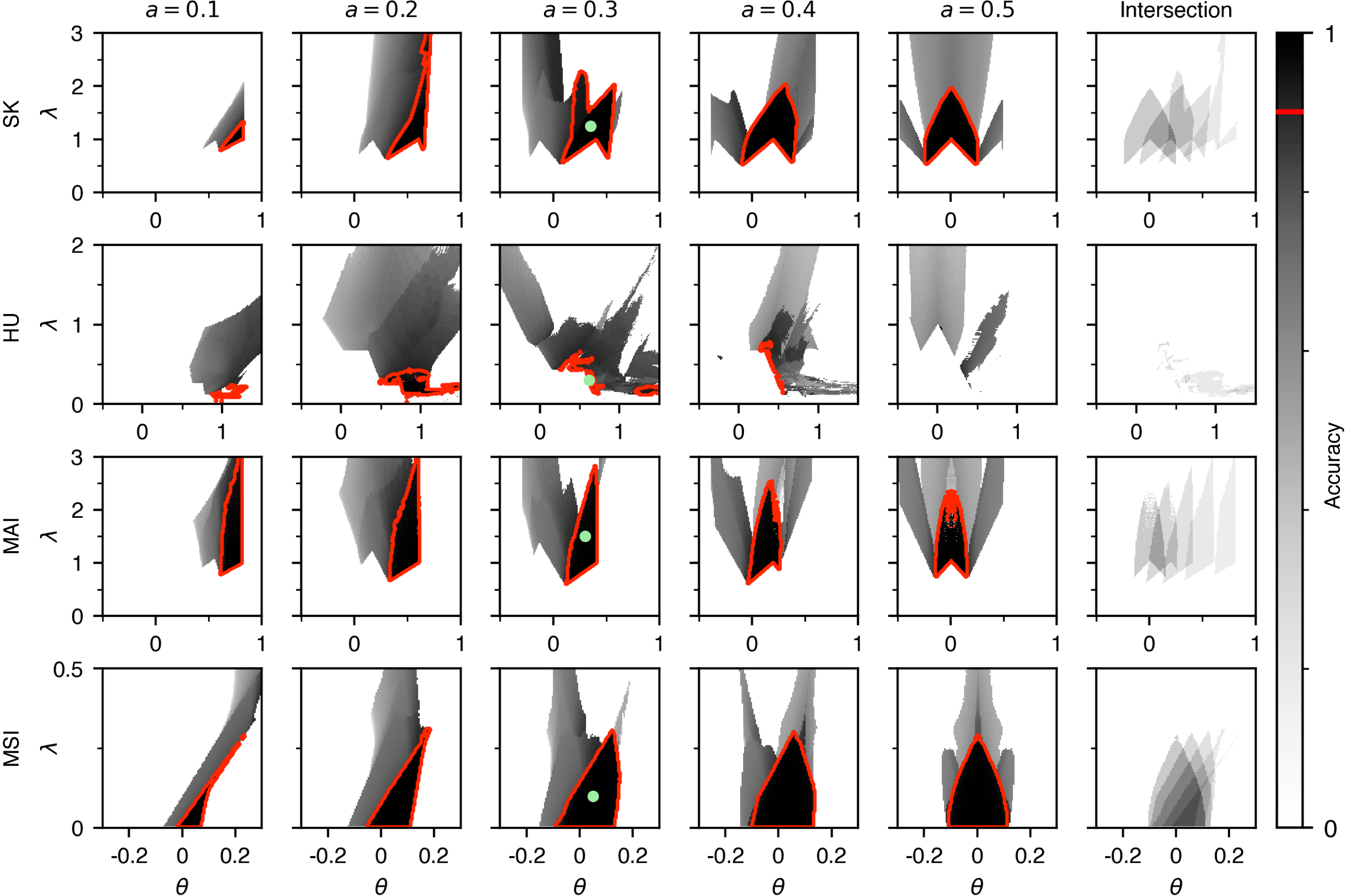}
    \caption{{\bf Parameter space of sequential retrieval.} The performance of each model for retrieving a cyclic sequence of four orthogonal patterns is evaluated at different combinations of the bias $\lambda$ and threshold $\theta$. These parameters were numerically sweeped with increments $\Delta\lambda=\Delta\theta=0.025$. The grayscale color represents the accuracy score (see Appendix~\ref{sec:computational-details}); the red contours represent regions of high accuracy ($> 0.9$). The first five columns correspond to different values of the pattern activity $a$, whereas the last column shows the overlay of the high-accuracy regions for these different activity levels. The green dots in the $a=0.3$ column correspond to the time series shown in Fig.~\ref{fig:examples}.}
    \label{fig:seqdyn-phasespace}%
\end{figure*}

Fig. \ref{fig:seqdyn-phasespace} presents $(\lambda, \theta)$ plots for different levels of activity $a$ in each of the four models. Shaded regions correspond to parameter combinations that produce sequential dynamics, and regions within the red contour correspond to high accuracy (above 0.9). It can be seen that both input and interaction modulation models have compact regions of parameter space that allow sequential dynamics. However, the HU model is the least robust compared to others, as small parameter changes lead to dysfunctional behavior, such as dynamics with very high frequencies or pattern dependent amplitudes and frequencies (see Appendix~\ref{sec:computational-details}).
 
To better visualize the dependence of model performance on the activity level $a$, we overlaid the regions of accurate retrieval (red contours in Fig.~\ref{fig:seqdyn-phasespace}) from different $a$ values, as shown in the last column of Fig.~\ref{fig:seqdyn-phasespace}. The HU model has very small retrieval regions for any $a$. For SK and MAI the retrieval region drifts from a positive value of $\theta$ towards 0 as $a$ increases. Notably, for MSI there is a compact region where the retrieval regions for different values of $a$ overlap. This suggests that MSI is able to retrieve sequences among patterns with varying activities, which we explore below.

\section{Variability in pattern activity} \label{sec:variable-patterns}

\begin{figure*}
    \centering
    \includegraphics[
    width=\textwidth
    ]{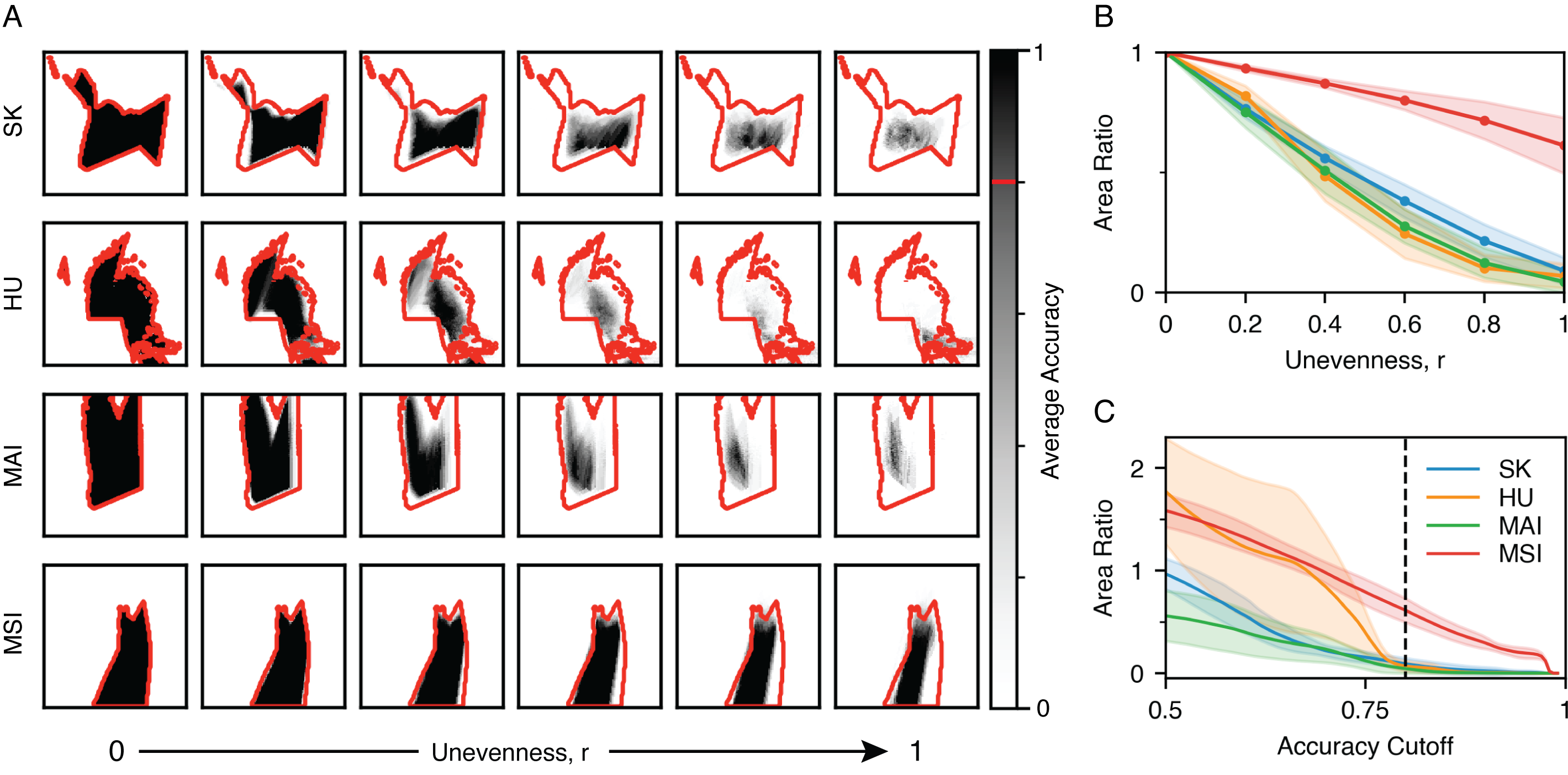}
    \caption{{\bf Retrieval for variable activities and sequence compositions.} {\bf A}: Phase diagrams are computed for each model at different unevenness of activity levels ($r$ values as marked in panel B). The parameter ranges are the same as for Fig.~\ref{fig:seqdyn-phasespace}. The retrieval accuracy at each point $(\theta, \lambda)$ is first binarized according to a cutoff of 0.8, and then averaged over all permutations. This average accuracy is colored using a grayscale, and the red contours represent the shaded area for $r=0$. {\bf B}: The ratio of the gray area to the red contoured region is calculated for each unevenness value $r$. The curves show the mean of the area ratios over sequence permutations (for an accuracy cutoff of 0.8), and the color shades show the standard deviations. {\bf C}: The area ratio for unevenness $r = 1$ is calculated as the accuracy cutoff is varied. The dashed line corresponds to a cutoff of 0.8 used in panel A \& B, above which all models fail except for MSI.}
    \label{fig:variable-activities}
\end{figure*}

So far we have focused on sequential retrieval of patterns with the same activity. To study how well the models can retrieve patterns with different activities, we consider patterns $\xi^\mu$ which each have a particular activity level $a_\mu$. We chose five patterns with $a_\mu$ equally spaced within the range $0.3 \pm 0.2 r$, where the ``unevenness'' parameter $r$ is varied between $0$ and $1$. Because the patterns have different activities, the order of these patterns in the sequence can affect retrieval. Therefore, we computed the mean accuracy of retrieval over all possible permutations of a given set of patterns (the scores are first binarized according to a cutoff and then averaged).

Fig.~\ref{fig:variable-activities}A shows the average accuracy as we vary the unevenness. The ratio of the gray area to the retrieval region for uniform patterns (red contour) represents the ability of each model to retrieve uneven patterns. It can be seen that the MSI model is robust to unevenness in pattern activity, as expected from our observation of Fig.~\ref{fig:seqdyn-phasespace}, as well as to the ordering of the patterns. We quantified this robustness in Fig.~\ref{fig:variable-activities}B, where the area of the retrieval region is plotted against the unevenness. While this measurement of robustness decays rapidly for other models, it is much more stable for MSI. Our results are not qualitatively affected by altering the accuracy cutoff, as shown in Fig.~\ref{fig:variable-activities}C.

\section{Dynamic storage capacity} \label{sec:storage-capacity}

So far we have studied  sequential retrieval of a small number of patterns. We now study how these models differ in their dynamic capacity, i.e., the ability to retrieve increasingly long sequences of patterns. We quantify the dynamic capacity $\alpha_D$ as the longest sequence of patterns which may be stored by a network of size $N$. The patterns are randomly generated with the same activity level ($a = 0.3$), and we average the accuracy over 100 sequence realizations.

\begin{figure*}
    \centering
    \includegraphics[
    width=\textwidth
    ]{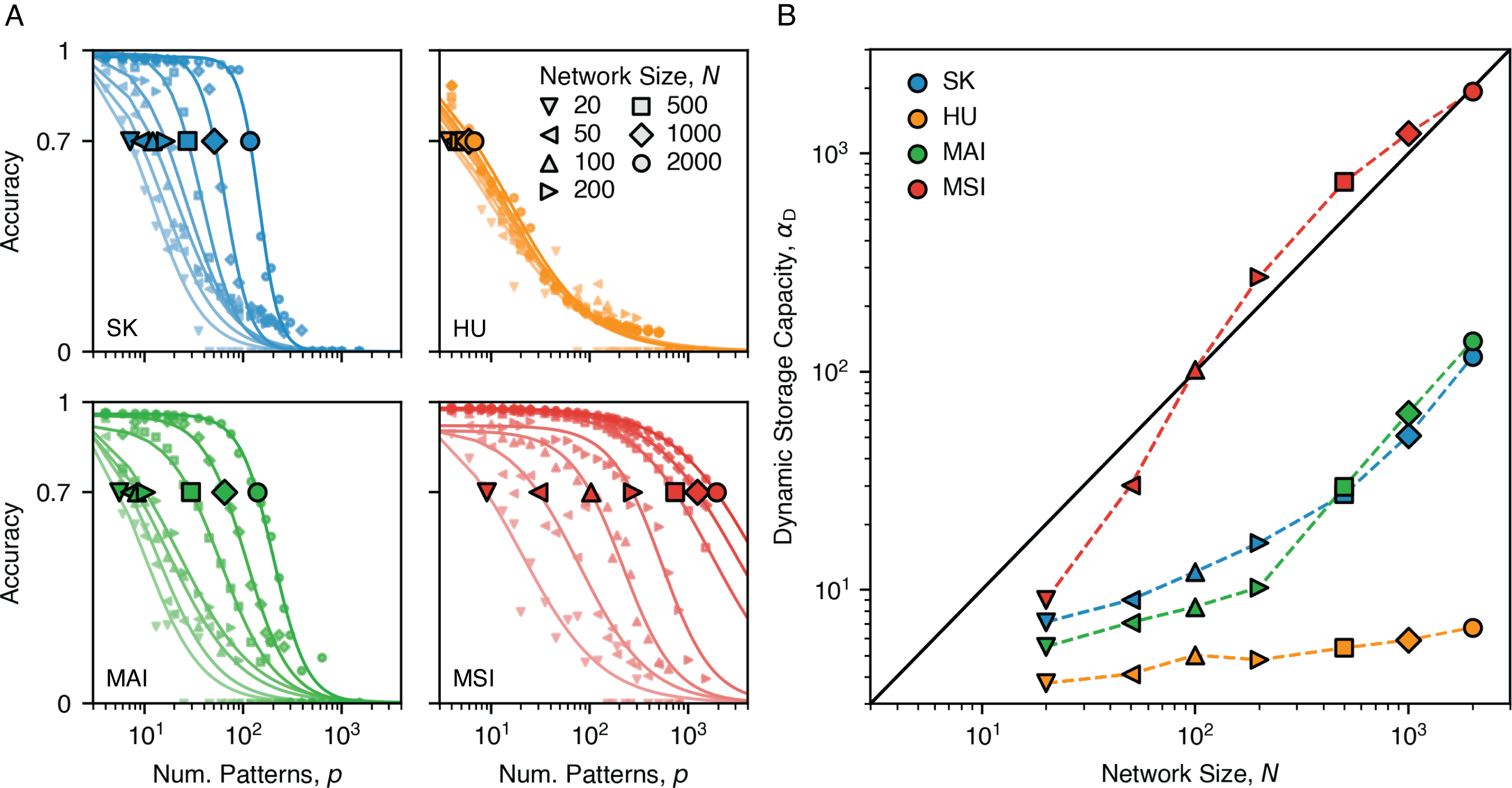}
    \caption{{\bf Dynamic capacity of sequential retrieval.} {\bf A}: For each network size $N$, the average accuracy of retrieval is plotted against the number of patterns $p$. The accuracy is calculated for a sequence of $p$ randomly generated patterns with $a=0.3$, averaged over 100 realizations. A sigmoidal curve is fitted for each $N$, and the dynamic storage capacity $\alpha_{\mathrm{D}}$ is defined as the value of $p$ where the average accuracy falls below 0.7, indicated by the marker. {\bf B}: The estimated values of $\alpha_{\mathrm{D}}$ are plotted against the size of the network $N$, showing different scaling for each model. The parameters used to simulate each model are the same as for Fig.~\ref{fig:examples} (see caption).
    }
    \label{fig:dynamic-capacity}
\end{figure*}

Fig.~\ref{fig:dynamic-capacity}A shows the accuracy as a function of the number of patterns for the four models in Box.~1. As the length of the sequence increases, the retrieval accuracy decreases. However, substantial differences exist in the behavior of these models. For instance, the HU model shows no apparent improvement in its capacity to store longer sequences as the network size increases. In contrast, the MSI model remains highly accurate for a large number of patterns. To corroborate this observation, we plot the dynamic storage capacity against the network size $N$, as shown in Fig.~\ref{fig:dynamic-capacity}B. While the range of $N$ is not sufficient to constitute a comprehensive scaling analysis, this figure shows that the MSI model vastly outperforms the other models in terms of dynamic storage capacity.

\section{Discussion}

In this work we have used the Hopfield network as a setup and formulated multiple models capable of sequentially retrieving stored patterns. In our framework, the transition between subsequent patterns is controlled by a set of feedback units, where feedback is coupled either directly to the main units in the case of input modulation, or to the interactions between these units in the case of interaction modulation.

While all models that we studied are capable of sequential retrieval, we showed that an interaction modulation model (MSI) outperforms all others. The MSI model is robust to variations in the activity level of the patterns and the ordering of the sequence, and is capable of retrieving much longer sequences than the others. The unique features of the MSI model are likely due to the dynamics of the network happening across a less rugged energy landscape (Box 1). In associative memory networks spurious minima arise as linear combinations of stored patterns, which result in a rugged landscape and dynamics which may be trapped in spurious minima. The MSI model, however, stores only a single pattern in $J_{ij}$ at any moment during sequential retrieval. We therefore expect that dynamics unfold on a smoother landscape, resulting in more accurate retrieval as observed.

Many models of sequential retrieval, such as those presented in Section~\ref{sec:interaction-input}, originate from the study of associative memory and the dynamics of real neurons. For example, in the SK model the role of feedback is played by slow asymmetric interactions \cite{sompolinsky1986temporal}. Later models, such as HU, introduced such slow feedback as neural fatigue, modeled as a neuron-specific threshold that increases when the neuron remains active \cite{horn1989neural}. Another model produces so-called latching dynamics through combining a slow time-dependent threshold with neurons of many internal states \cite{russo2012cortical}.

Coupling feedback to interactions has been studied in various forms within the context of neural networks. In \cite{dehaene1987neural}, the authors are motivated by a type of allostery among neurons -- regulation of the efficacy of a given synapse by the activity of another synapse. Their model allows pairwise interactions to be modified by other neurons in the network, and bears resemblance to our MAI model. This phenomenon, by which a synapse that is not currently active can be strengthened or weakened by the firing of a third modulatory neuron, is referred to as heterosynaptic plasticity \cite{Bailey2000, Shouval2002, Wang2006, Chater2021}. It has inspired network models that can learn sequences through synaptic competition \cite{fiete2010spike}. Other biological phenomena, such as neural fatigue \cite{Markram1996}, have also inspired models that modulate neural interactions rather than individual thresholds. For example, in   \cite{pantic2002associative} a slow feedback is introduced to depress the synapses between neurons. This model can be reformulated in our framework of interaction modulation similarly to the MSI model, as presented in Appendix~\ref{sec:pantic-like}. In \cite{Recanatesi2017}, synaptic depression is used to control transitions between patterns by externally modulating a global inhibition of all interactions among neurons, which allows for transitions between correlated patterns. Instead of controlling such inhibition externally, we can modify their model to couple the inhibition to a slow feedback and reformulate it within our framework, as presented in Appendix~\ref{sec:tsodyks-like}.

Separate from a biological setting, the mechanism that we introduce as feedback appears widely within control theory, where it is typically studied in the context of control through linear input modulation \cite{bechhoefer2021control, Liu2016}. However, interaction modulation is a less studied form of controlling network dynamics \cite{nepusz2012controlling}, especially when the interactions are structured like in the Hopfield network. Our results suggest that interaction modulation may be a new paradigm for controlling nonlinear dynamics.

\section{Conclusion}

In biology, interactions among the components of a system are often effective, coarse-grained descriptions of complicated microscopic mechanisms. The strength of such effective interactions can be tuned by modifying the underlying mechanisms. Modification of interaction strengths is quite common among biological systems, such as allosteric regulation of protein interactions \cite{phillips2020molecular} and trait-mediated modification of species interactions \cite{Wootton1994, Terry2017}. Yet, the benefits of being able to modify interactions is under-explored theoretically. Here we have demonstrated that interaction modulation can be an effective way of controlling the stability of system configurations and the direction of its dynamics, which may be important for biological functions and evolution.

In complex systems, the interactions among many constituent units give rise to various collective behaviors, such as coherent motion \cite{vicsek1995novel, toner1998flocks, vicsek2012collective} or collaborative functions \cite{Hashimura2019}. The sequential transition of the system between multiple metastable configurations that we modeled here is one type of collective behavior, and we have shown that interaction modulation is a robust way of controlling such behavior. Our study may inspire future work on exploring the role of interaction modulation in other situations.

\bibliographystyle{aip}
\bibliography{neural_dyn}

\begin{thebibliography}{10}

\bibitem{wallenstein1998hippocampus}
G.~V. Wallenstein, M.~E. Hasselmo, and H.~Eichenbaum,
\newblock Trends in neurosciences {\bf 21}, 317 (1998).

\bibitem{Eichenbaum_2013}
H.~Eichenbaum,
\newblock Trends in Cognitive Sciences {\bf 17}, 81 (2013).

\bibitem{wahl2009spliceosome}
M.~C. Wahl, C.~L. Will, and R.~L{\"u}hrmann,
\newblock cell {\bf 136}, 701 (2009).

\bibitem{Datta2016}
M.~S. Datta, E.~Sliwerska, J.~Gore, M.~F. Polz, and O.~X. Cordero,
\newblock Nature Communications {\bf 7} (2016).

\bibitem{Liu2016}
Y.-Y. Liu and A.-L. Barab{\'{a}}si,
\newblock Reviews of Modern Physics {\bf 88} (2016).

\bibitem{Wodak2019}
S.~J. Wodak et~al.,
\newblock Structure {\bf 27}, 566 (2019).

\bibitem{Bailey2000}
C.~H. Bailey, M.~Giustetto, Y.-Y. Huang, R.~D. Hawkins, and E.~R. Kandel,
\newblock Nature Reviews Neuroscience {\bf 1}, 11 (2000).

\bibitem{Hopfield1982}
J.~J. Hopfield,
\newblock Proceedings of the National Academy of Sciences {\bf 79}, 2554
  (1982).

\bibitem{de2012reaction}
A.~De~Martino, D.~De~Martino, R.~Mulet, and G.~Uguzzoni,
\newblock PloS one {\bf 7}, e39849 (2012).

\bibitem{sartori2020lessons}
P.~Sartori and S.~Leibler,
\newblock Proceedings of the National Academy of Sciences {\bf 117}, 114
  (2020).

\bibitem{Power2015}
D.~A. Power et~al.,
\newblock Biology Direct {\bf 10}, 69 (2015).

\bibitem{sompolinsky1986temporal}
H.~Sompolinsky and I.~Kanter,
\newblock Physical review letters {\bf 57}, 2861 (1986).

\bibitem{horn1989neural}
D.~Horn and M.~Usher,
\newblock Physical Review A {\bf 40}, 1036 (1989).

\bibitem{dehaene1987neural}
S.~Dehaene, J.-P. Changeux, and J.-P. Nadal,
\newblock Proceedings of the National Academy of Sciences {\bf 84}, 2727
  (1987).

\bibitem{buhmann1987noise}
J.~Buhmann and K.~Schulten,
\newblock EPL (Europhysics Letters) {\bf 4}, 1205 (1987).

\bibitem{kleinfeld1988associative}
D.~Kleinfeld and H.~Sompolinsky,
\newblock Biophysical Journal {\bf 54}, 1039 (1988).

\bibitem{amit1988neural}
D.~J. Amit,
\newblock Proceedings of the National Academy of Sciences {\bf 85}, 2141
  (1988).

\bibitem{russo2012cortical}
E.~Russo and A.~Treves,
\newblock Physical Review E {\bf 85}, 051920 (2012).

\bibitem{naim2020fundamental}
M.~Naim, M.~Katkov, S.~Romani, and M.~Tsodyks,
\newblock Physical review letters {\bf 124}, 018101 (2020).

\bibitem{osat2022non}
S.~Osat and R.~Golestanian,
\newblock arXiv preprint arXiv:2201.10362  (2022).

\bibitem{sompolinsky1988chaos}
H.~Sompolinsky, A.~Crisanti, and H.-J. Sommers,
\newblock Physical review letters {\bf 61}, 259 (1988).

\bibitem{Shouval2002}
H.~Z. Shouval, G.~C. Castellani, B.~S. Blais, L.~C. Yeung, and L.~N. Cooper,
\newblock Biological Cybernetics {\bf 87}, 383 (2002).

\bibitem{Wang2006}
Y.~Wang et~al.,
\newblock Nature Neuroscience {\bf 9}, 534 (2006).

\bibitem{Chater2021}
T.~E. Chater and Y.~Goda,
\newblock Current Opinion in Neurobiology {\bf 67}, 106 (2021).

\bibitem{fiete2010spike}
I.~R. Fiete, W.~Senn, C.~Z. Wang, and R.~H. Hahnloser,
\newblock Neuron {\bf 65}, 563 (2010).

\bibitem{Markram1996}
H.~Markram and M.~Tsodyks,
\newblock Nature {\bf 382}, 807 (1996).

\bibitem{pantic2002associative}
L.~Pantic, J.~J. Torres, H.~J. Kappen, and S.~C. Gielen,
\newblock Neural Computation {\bf 14}, 2903 (2002).

\bibitem{Recanatesi2017}
S.~Recanatesi, M.~Katkov, and M.~Tsodyks,
\newblock Neural Computation {\bf 29}, 2684 (2017).

\bibitem{bechhoefer2021control}
J.~Bechhoefer,
\newblock {\em Control Theory for Physicists},
\newblock Cambridge University Press, 2021.

\bibitem{nepusz2012controlling}
T.~Nepusz and T.~Vicsek,
\newblock Nature Physics {\bf 8}, 568 (2012).

\bibitem{phillips2020molecular}
R.~Phillips,
\newblock The molecular switch,
\newblock in {\em The Molecular Switch}, Princeton University Press, 2020.

\bibitem{Wootton1994}
J.~T. Wootton,
\newblock Annual Review of Ecology and Systematics {\bf 25}, 443 (1994).

\bibitem{Terry2017}
J.~C.~D. Terry, R.~J. Morris, and M.~B. Bonsall,
\newblock Ecology Letters {\bf 20}, 1219 (2017).

\bibitem{vicsek1995novel}
T.~Vicsek, A.~Czir{\'o}k, E.~Ben-Jacob, I.~Cohen, and O.~Shochet,
\newblock Physical review letters {\bf 75}, 1226 (1995).

\bibitem{toner1998flocks}
J.~Toner and Y.~Tu,
\newblock Physical review E {\bf 58}, 4828 (1998).

\bibitem{vicsek2012collective}
T.~Vicsek and A.~Zafeiris,
\newblock Physics reports {\bf 517}, 71 (2012).

\bibitem{Hashimura2019}
H.~Hashimura, Y.~V. Morimoto, M.~Yasui, and M.~Ueda,
\newblock Communications Biology {\bf 2} (2019).

\end{thebibliography}

\clearpage

\appendix
\section{Other models of interaction modulation} \label{sec:alt-models}

Sequential dynamics through interaction modulation can be implemented in models other than those presented in the main text. Here we provide two more examples.

\subsection{Complement of MSI} \label{sec:pantic-like}
In the presented MSI model the symmetric interactions $J_{ij}$ have dynamics such that at any time only one pattern is accessible, as only one $c_\mu$ is active. The network retrieves the pattern and, after some time, this pattern is purged from $J_{ij}$ and a new pattern is stabilized and retrieved. Here, we present a complementary model where at any given time all patterns are present in $J_{ij}$ except one. When the network retrieves one pattern, the corresponding $c_\mu$ will slowly suppress it in $J_{ij}$ while all other patterns remain. This model can be described similarly to MSI in Box~1, except that
\begin{align}
    J_{ij}(c_\mu) = \frac{1}{N} \sum_\mu (\xi_i^{\mu} - a) (\xi_j^{\mu} - a) (1 - c_\mu) \;.
\end{align}
When $c_\mu$ increases and $\xi^\mu$ is no longer stable, the network will be pushed towards the next pattern $\xi^{\mu+1}$ by the asymmetric interactions $\tilde{J}_{ij}$ as in MSI. This model is similar in spirit to that studied in \cite{pantic2002associative}, where part of the symmetric interactions is depressed after a pattern is retrieved for some time. Example dynamics for the complement of MSI are shown in Fig. \ref{fig:alt-examples}A.

\subsection{Global inhibition} \label{sec:tsodyks-like}

\begin{figure}%
    \centering
    \includegraphics[
    ]{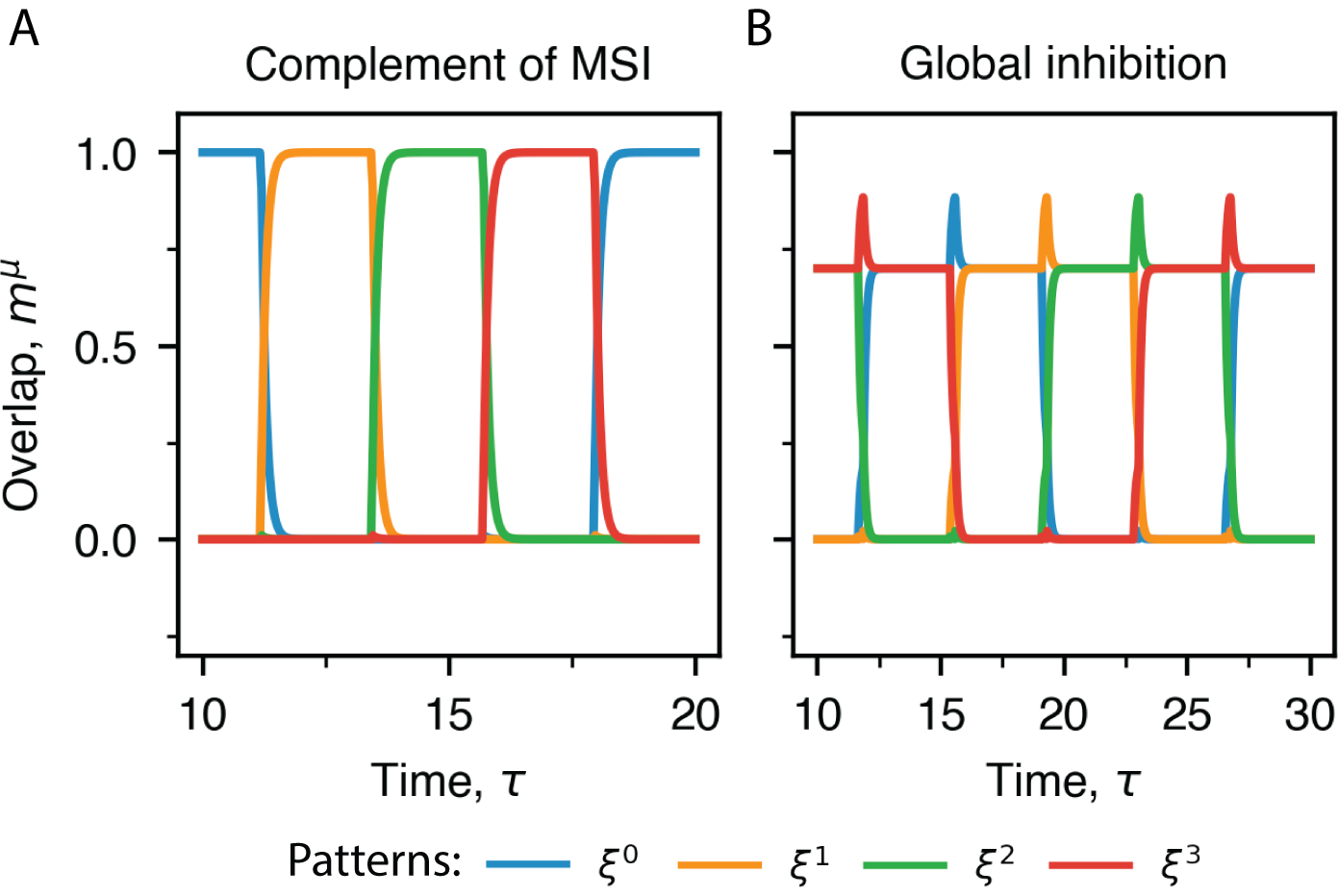}
  \caption{{\bf Example dynamics for alternative models.} {\bf A}: Complement of MSI model ($\lambda=0.05, \theta=0, a=0.3$) {\bf B}: Modified global inhibition model $(\lambda=0.7, \theta=0, a=0.3)$. Both models are described in Appendix~\ref{sec:alt-models} and simulated for a cyclic sequence of four orthogonal patterns.}
    \label{fig:alt-examples}%
\end{figure}

Another form of interaction modulation has been considered in \cite{Recanatesi2017}, where the symmetric interactions take the form
\begin{equation} \label{eq:tsodyks-like}
 J_{ij} = \frac{1}{N}\left( \sum_\mu (\xi_i^\mu - a) (\xi_j^\mu - a) - \phi \right)\quad.
\end{equation}
In this equation, the parameter $\phi$ represents a global inhibition of all symmetric interactions. This parameter is externally controlled to oscillate between a minimum value at which patterns can be retrieved and a maximum value at which only part of the previously retrieved pattern can remain active. When $\phi$ is reduced again, the next pattern retrieved will be one that happens to have the largest intersection with the active part \cite{Recanatesi2017}. We can modify the model so that the dynamics of $\phi$ is autonomous through feedback, rather than being modulated externally. For example, we can have
\begin{equation}
    \phi = \theta \sum_\mu c_\mu\quad,
\end{equation}
where the parameter $\theta$ controls the maximum inhibition strength. Example dynamics for the modified global inhibition model are shown in Fig.~\ref{fig:alt-examples}B.

\section{Computing retrieval accuracy} \label{sec:computational-details}

 To evaluate the performance of sequential retrieval over an extended period of time, we introduce a scoring scheme that first calculates instantaneous scores and then averages them over time to produce an overall accuracy. The instantaneous score function is defined for each pattern as:
\begin{equation} \label{eq:score-instantaneous}
 s^\mu(\{m^\nu\}) = \frac{G(m^\mu)}{\sum_\nu G(m^\nu) + \epsilon}
\end{equation}
where
\begin{equation} \label{eq:expit}
 G(m^\mu) = \frac{ \operatorname{expit}(m^\mu; \beta, \theta) - \operatorname{expit}(-1; \beta, \theta)}{\operatorname{expit}(1; \beta, \theta) - \operatorname{expit}(-1; \beta, \theta)}
 \quad \text{and} \quad
 \operatorname{expit}(x; \beta, \theta) = \frac{1}{1 + \mathrm{e}^{-\beta(x - \theta)}}.
\end{equation}
Our construction of $G(m^\mu)$ attenuates $m^\mu$ below some threshold $\theta$ towards 0 and amplifies $m^\mu$ above $\theta$ towards 1, so instances of retrieval correspond to a single high $s^\mu$ when the patterns are orthogonal. The parameters were chosen as $\beta=10$ and $\theta=1-a$ for all analyses, and $\epsilon$ was chosen to be $10^{-5}$ to make the instantaneous score well defined even when $m^\mu = 0$ for all $\mu$.

Each pattern is retrieved and remains stable for a continuous interval of time, which we call an \textit{instance} of retrieval. We identify such intervals as blocks of time when $G(m^\mu) \approx 1$, which corresponds to $m^\mu > \theta$. The score of an instance of retrieval amounts to the time average of the instantaneous scores over the interval:
\begin{equation}
    \label{eq:instance-score}
    S^\mu = \frac{1}{t_2 - t_1} \int_{t_1}^{t_2} s^\mu(\{m^\nu\}) \, dt \,,
\end{equation}
where $t_1$ and $t_2$ are the bounds of the interval. The time series of network dynamics is typically composed of many retrieval instances, so we define the \emph{overall accuracy} of sequential retrieval as the average score over many retrieval instances in the time series. In all cases, the network is simulated with $\tau=10$ and $\Delta t=0.1$ for 6000 time steps. We compute the accuracy only for the latter half of each time series to avoid the transient dynamics in the beginning.

\begin{figure}%
    \centering
    \includegraphics[
    ]{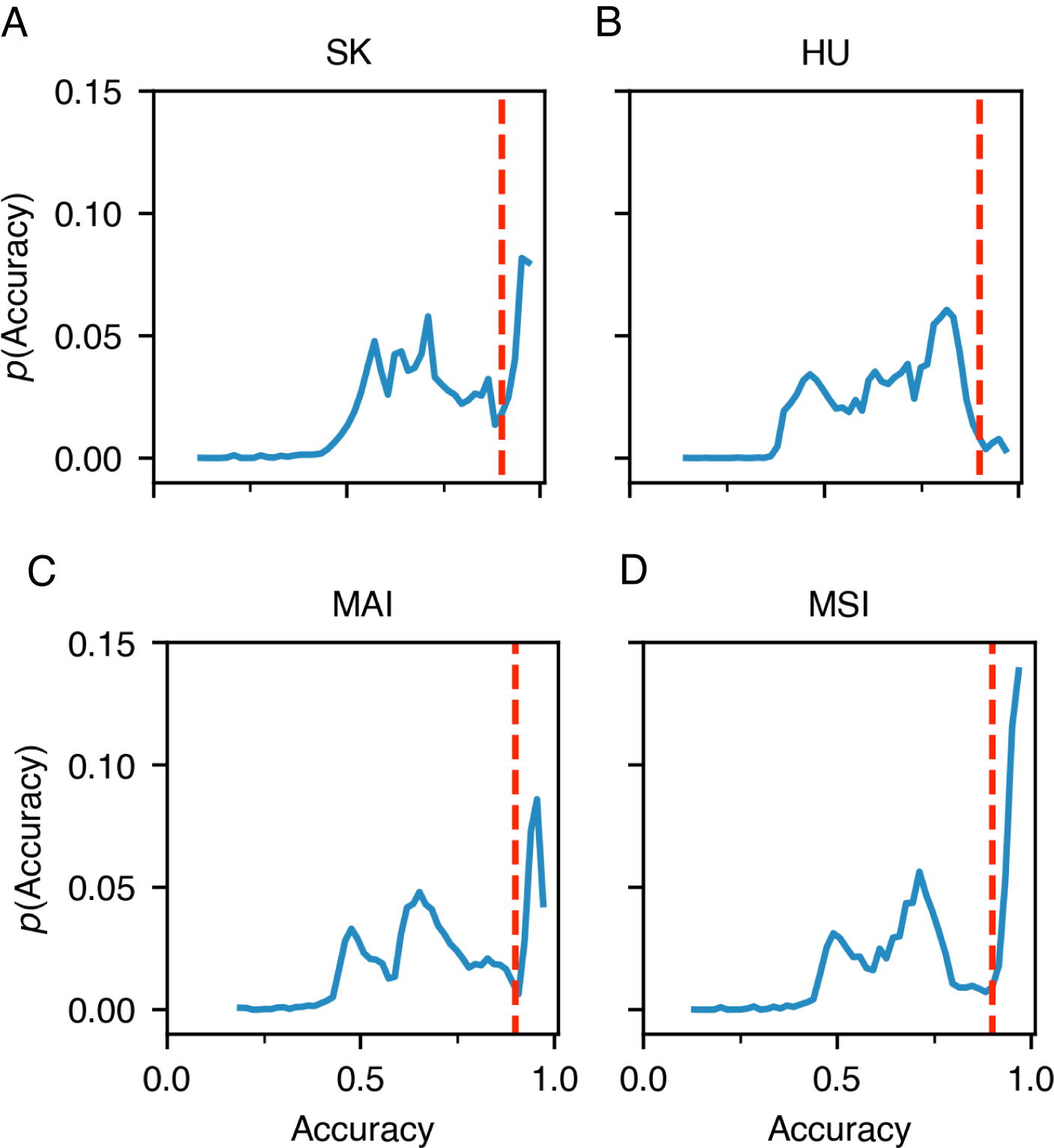}
  \caption{{\bf Accuracy distributions.} The distributions of accuracies over the parameter phase space in Fig. \ref{fig:seqdyn-phasespace} is calculated and marginalized over $a$ for ({\bf A-B}) input modulation and ({\bf C-D}) interaction modulation models. The rightmost peak, corresponding to accurate retrieval, is separated by an accuracy cutoff of 0.9 (only a small peak is present in HU). For clarity, scores of zero are omitted.}
    \label{fig:seqdyn-scorestats}%
\end{figure}

In the phase space analysis of Section \ref{sec:phase} we choose a cutoff between high-accuracy and low-accuracy regions of retrieval, indicated by the red contours in Fig. \ref{fig:seqdyn-phasespace}. To determine an appropriate accuracy cutoff, we examined the distribution of scores over parameter space ($\lambda, \theta, a)$ for each model, as shown in Fig. \ref{fig:seqdyn-scorestats}. The right-most peak (corresponding to high-accuracy retrieval) is separated from the remaining peaks by a cutoff accuracy of 0.9.

\begin{figure}
    \centering
    \includegraphics[
    width=\textwidth
    ]{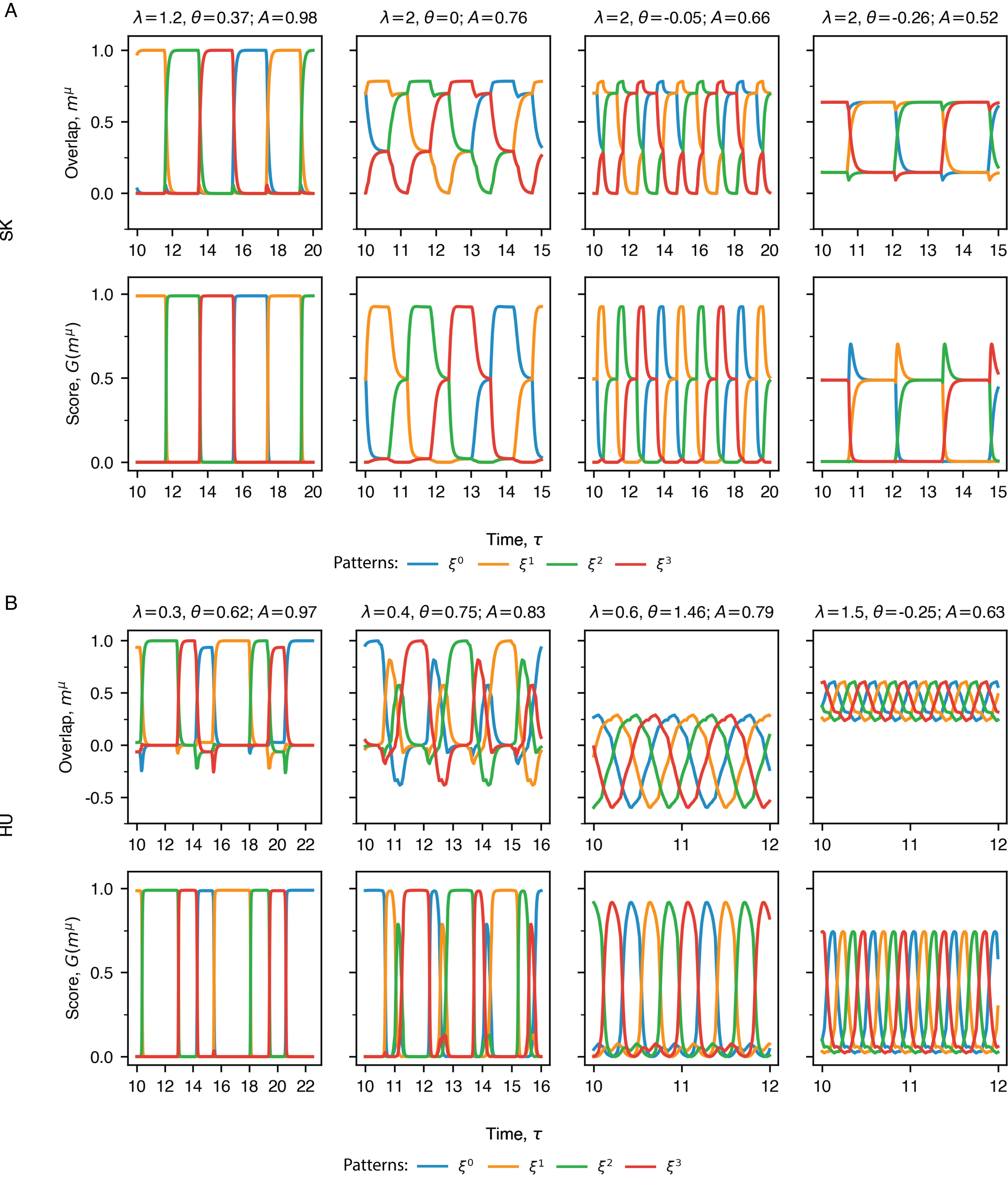}
  \caption{{\bf Representative dynamics for input modulation.} {\bf A}: SK model. {\bf B}: HU model. The pattern activity is 0.3 for all panels; the parameters and the resulting accuracy are listed above each column. The first row corresponds to the time series of overlaps $m^\mu$, while the second row corresponds to the instantaneous score $G(m^\mu)$. The parameters for $G(\cdot)$ are the same as those used to generate Fig. \ref{fig:seqdyn-phasespace}.}
    \label{fig:input-modulation-time-series}%
\end{figure}

\begin{figure}%
    \centering
    \includegraphics[
    width=\textwidth
    ]{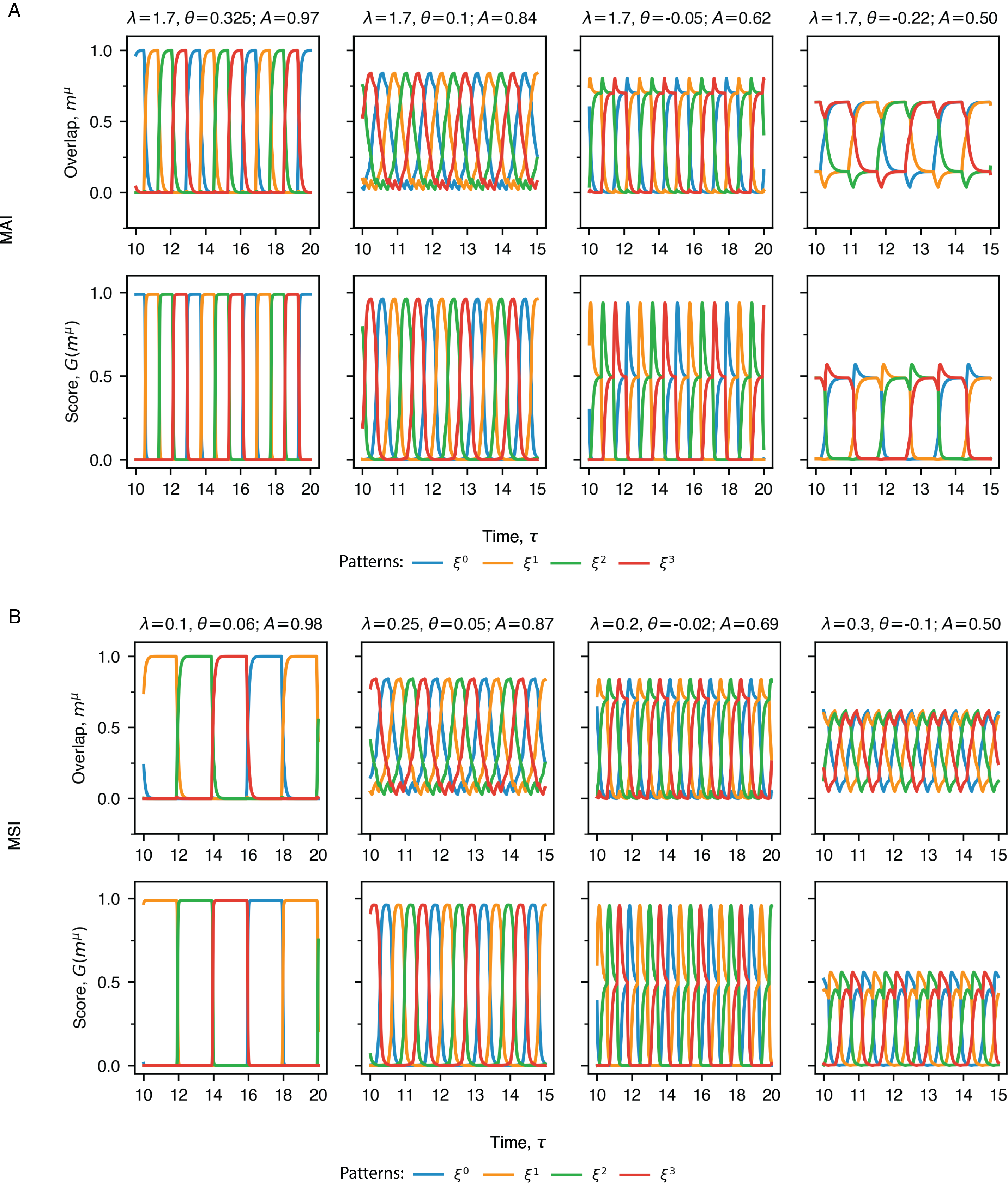}
    \caption{{\bf Representative dynamics for interaction modulation.} {\bf A}: MAI model. {\bf B}: MSI model. The pattern activity is 0.3 for all panels; the parameters and the resulting accuracy are listed above each column. The first row corresponds to the time series of overlaps $m^\mu$, while the second row corresponds to the instantaneous score $G(m^\mu)$. The parameters for $G(\cdot)$ are the same as those used to generate Fig. \ref{fig:seqdyn-phasespace}.}
    \label{fig:interaction-modulation-time-series}%
\end{figure}

\end{document}